\documentclass[fleqn,usenatbib]{mnras}

\usepackage{newtxtext,newtxmath}
\usepackage[T1]{fontenc}

\DeclareRobustCommand{\VAN}[3]{#2}
\let\VANthebibliography\thebibliography
\def\thebibliography{\DeclareRobustCommand{\VAN}[3]{##3}\VANthebibliography}

\usepackage{graphicx}	% Including figure files
\usepackage{amsmath}	% Advanced maths commands
\usepackage{float}
\usepackage[utf8]{inputenc}
\usepackage{babel}
\usepackage{hyperref}
\usepackage{dblfloatfix}
\usepackage{float}

\usepackage{color, hyperref}
\usepackage{xcolor}
\usepackage[normalem]{ulem}
\usepackage{CJK}
\usepackage{threeparttable}
\usepackage{nicefrac}
\usepackage{deluxetable}
\usepackage{longtable}

\newbox\grsign \setbox\grsign=\hbox{$>$} \newdimen\grdimen \grdimen=\ht\grsign
\newbox\simlessbox \newbox\simgreatbox
\setbox\simgreatbox=\hbox{\raise.5ex\hbox{$>$}\llap
     {\lower.5ex\hbox{$\sim$}}}\ht1=\grdimen\dp1=0pt
\setbox\simlessbox=\hbox{\raise.5ex\hbox{$<$}\llap
     {\lower.5ex\hbox{$\sim$}}}\ht2=\grdimen\dp2=0pt
\def\simgreater{\mathrel{\copy\simgreatbox}}
\def\simless{\mathrel{\copy\simlessbox}}
\newbox\simppropto
\setbox\simppropto=\hbox{\raise.5ex\hbox{$\sim$}\llap
     {\lower.5ex\hbox{$\propto$}}}\ht2=\grdimen\dp2=0pt

\newcommand{\Msun}{\,M_{\odot}}

\usepackage{graphicx}
\DeclareGraphicsExtensions{.pdf,.pdf}
\graphicspath{{./}{figures/}}

\usepackage{graphicx}
\DeclareGraphicsExtensions{.pdf,.pdf}
\graphicspath{{./}{figures/}}

\title[Chemistry of GE/S vs Milky Way Satellites]{A Comparative Analysis of the Chemical Compositions of Gaia-Enceladus/Sausage and Milky Way Satellites using APOGEE}

\author[L. Fernandes et al.]{Laura Fernandes$^{1}$,\thanks{E-mail: Lfernandes2@msn.com (LF); R.P.Schiavon@ljmu.ac.uk (RPS)}
Andrew C. Mason$^{1}$,
Danny Horta$^{1}$,
Ricardo P. Schiavon$^{1}$,
Christian Hayes$^{2}$, 
\newauthor
Sten Hasselquist$^{3}$,
Diane Feuillet$^{4}$,
Rachael L. Beaton$^{5,6}$,
Henrik Jönsson$^{7}$,
Shobhit Kisku$^{1}$,
\newauthor
Ivan Lacerna$^{8,9}$,
Jianhui Lian$^{10}$,
Dante Minniti$^{11,12}$,
Sandro Villanova$^{13}$
\\
$^{1}$ Astrophysics Research Institute, Liverpool John Moores University, 146 Brownlow Hill, Liverpool L3 5RF, UK\\
$^{2}$ NRC Herzberg Astronomy and Astrophysics, 5071 West Saanich Road, Victoria, B.C., Canada, V9E 2E7 \\
$^{3}$ Space Telescope Science Institute, 3700 San Martin Drive, Baltimore, MD 21218, USA\\
$^{4}$ Lund Observatory, Department of Astronomy and Theoretical Physics, Box 43, SE-221 00 Lund, Sweden\\
$^{5}$ Department of Astrophysical Sciences, 4 Ivy Lane, Princeton University, Princeton, NJ 08544\\
$^{6}$ The Observatories of the Carnegie Institution for Science, 813 Santa Barbara St., Pasadena, CA~91101\\
$^{7}$ Materials Science and Applied Mathematics, Malmö University, SE-205 06 Malmö, Sweden\\
$^{8}$ Instituto de Astronomía y Ciencias Planetarias, Universidad de Atacama, Copayapu 485, Copiap\'o, Chile\\
$^{9}$ Millennium Institute of Astrophysics, Nuncio Monsenor Sotero Sanz 100, Of. 104, Providencia, Santiago, Chile \\
$^{10}$ Department of Physics and Astronomy, University of Utah,
115 S. 1400 E., Salt Lake City, UT 84112, USA \\
$^{11}$ Departamento de Ciencias Físicas, Facultad de Ciencias Exactas, Universidad Andres Bello, Fernández Concha 700, Las Condes, Santiago, Chile \\
$^{12}$ Vatican Observatory, Vatican City State, V-00120, Italy\\
$^{13}$ Departamento de Astronomía, Universidad de Concepci\'on, Casilla 160-C, Concepci\'on, Chile\\
}

\date{Accepted XXX. Received YYY; in original form ZZZ}

\pubyear{2021}

\DeclareUnicodeCharacter{2212}{-}

\begin{document}
\label{firstpage}
\pagerange{\pageref{firstpage}--\pageref{lastpage}}
\maketitle

\begin{abstract}
We use data from the 17\textsuperscript{th} data release of the Apache Point Observatory Galactic Evolution Experiment (APOGEE 2) to contrast the chemical composition of the recently discovered Gaia Enceladus/Sausage system (GE/S) to those of ten Milky Way (MW) dwarf satellite galaxies: LMC, SMC, Bo\"otes~I, Carina, Draco, Fornax, Sagittarius, Sculptor, Sextans and Ursa Minor.  Our main focus is on the distributions of the stellar populations of those systems in the [Mg/Fe]-[Fe/H] and [Mg/Mn]-[Al/Fe] planes, which are commonly employed in the literature for chemical diagnosis and where dwarf galaxies can be distinguished from {\it in situ} populations.  We show that, unlike MW satellites, a GE/S sample defined purely on the basis of orbital parameters falls almost entirely within the  locus of ``accreted'' stellar populations in chemical space, which is likely caused by an early quenching of star formation in GE/S.  Due to a more protracted history of star formation, stars in the metal-rich end of the MW satellite populations are characterized by lower [Mg/Mn] than those of their GE/S counterparts.  The chemical compositions of GE/S stars are consistent with a higher early star formation rate than MW satellites of comparable and even higher mass, suggesting that star formation in the early universe was strongly influenced by other parameters in addition to mass.  We find that the direction of the metallicity gradient in the [Mg/Mn]--[Al/Fe] plane of dwarf galaxies is an indicator of the early star formation rate of the system.

\end{abstract}


%%%%%%%%%%%%%%%%%%%%%%%%%%%%%%%%%%%%%%%%%%%%%%%%%%

%%%%%%%%%%%%%%%%% BODY OF PAPER %%%%%%%%%%%%%%%%%%

%****** FIGURE ******** 
%\begin{figure*}
%\includegraphics[width=0.8\textwidth]{Images/2D_plot.png}
%\caption{Two-dimensional abundance distribution of 733,901 stars in the 17\textsuperscript{th} data release (DR17) of the SDSS-IV/APOGEE-2. Represented in the chemical planes of [Mg/Fe] vs [Fe/H] (left panel) and [Mg/Mn] vs [Al/Fe] (right panel). Prominent features (high density regions) include the stellar populations in the high- and low- $\alpha$ thick and thin disks of the MW, the Magellanic Clouds and the accreted system, Gaia-Enceladus/Sausage (GE/S).  
%\label{fig:2dhist}}
%\end{figure*}

\section{Introduction}

Our current understanding of galaxy formation is immersed in the framework of the Lambda Cold Dark Matter ($\Lambda$CDM) model.  In this model, galaxies are an ensemble of stars, gas, and dust bound by massive dark matter haloes, built largely by the merging of smaller systems \citep[e.g.,][]{white1978core,Blumenthal1984,frenk2012dark}.  In this context, the Milky Way (MW) is of great importance, as the galaxy for which we can obtain the most detailed information.  Naturally, one should expect that our knowledge of the MW should be used to constrain galaxy formation theory. While most of the stellar mass in the MW is contained in its bar and both thin and thick disks (\citealt{morris1996galactic}; \citealt{barbuy2018chemodynamical}; \citealt{Nataf_2017}; \citealt{rich2013}; \citealt{beraldo_disk}), these structures are enveloped by a large stellar halo which contains about 1.3 × 10\textsuperscript{9} $\Msun$ in the form of stars (\citealt{MackerethBovy2020}).
 
%\smallskip 
%(\citealt{queiroz2020milky}; \citealt{van2011galaxy}; \citealt{palla2020chemical}).

%Most importantly, due to long dynamical timescales, the stellar halo contains important clues for the accretion history of the MW.

%\smallskip 

Galaxy mergers were very common in the early universe.  Local evidence of the merging activity of the MW has been accumulating over the years, with the identification of the remnants of multiple accretion events, starting with the discovery of the Sagittarius dwarf spheroidal (Sgr dSph) by \cite{ibata1994dwarf}.  In addition, various stellar streams have been identified as remnants of past accretion events \citep[e.g.,][]{Belokurov2006}.  Due to the stellar halo's long dynamical timescale, such relatively recent accretion events can be identified in the form of spatial substructure. However, discerning the remnants of earlier accretion requires additional information, in the form of, e.g., kinematics, chemical compositions, and ages \citep[e.g.,][]{Nissen&Schuster_2010, Nissen&Schuster11, Schuster_2012, 2018_Hayes}

With the advent of large spectroscopic surveys, we entered the golden age of Galactic archaeology. The information acquired by past, ongoing and future surveys such as RAVE, (\citealt{steinmetz2006radial}), SEGUE (\citealt{yanny2009segue}), LAMOST (\citealt{cui2012large}), GALAH (\citealt{de2015galah}), \textit{Gaia} mission (\citealt{gaia2016}), WEAVE (\citealt{2016ASPC..507...97D}), APOGEE (\citealt{majewski2017apache}), H3 (\citealt{conroy2019mapping}), MOONS  (\citealt{cirasuolo2011moons}; \citealt{gonzalez2020moons}), 4MOST (\citealt{4MOSTdejong}) is very quickly advancing our understanding of the history of the MW.  Chemical compositions and orbital information is becoming available for millions of stars in the MW and its Local Group companions.  In addition, with the launching of the Gaia satellite \citep{gaia2016}, detailed phase space information has become available, enabling the identification of further substructures in the stellar halo of the Galaxy, associated with various accretion events, including  Gaia-Enceladus/Sausage, a massive dwarf galaxy accreted to the MW about $\sim$10Gyr ago \citep{belokurov2018co,haywood2018disguise,helmi2018merger,mackereth2019origin}, as well as Heracles \citep{Horta2021evidence}, and various other substructures \cite[e.g.,][]{koppelman2019multiple,belokurov2018co,kruijssen2020kraken,naidu2020evidence,Horta2022b}.

%\smallskip 

\cite{Hawkins_2015} and \cite{das2020ages} have recently proposed that a combination of the abundances of Mn, Mg, Al, and Fe can be used to chemically discriminate accreted populations from their counterparts formed {\it in situ}.  However, \cite{Horta2021evidence} used chemical evolution models to show that the region occupied by accreted populations in the [Mg/Mn] vs [Al/Fe] plane has a non-negligible presence of stars formed {\it in situ}.  

%It is thus important to examine the distribution of the stellar populations of accreted systems, such as Gaia-Enceladus, with those of accreted dwarf satellites of the Milky Way and {\it in situ} populations to clarify how they are individually distributed in that plane. 

It is difficult to disentangle metal-poor {\it in situ} populations from their accreted counterparts based on purely kinematic or orbital criteria.  To minimise inter-sample contamination, one has to often resort to additional criteria based on chemical compositions, introducing sample biases that prevent a clean analysis of the chemical compositions of accreted and {\it in situ} populations.  That difficulty can be overcome by resorting to data from external systems, on the assumption that the chemistry of their stellar populations is dominated by intrinsic evolutionary effects.  

\smallskip

In this paper we contrast the distribution of stars from GE/S system in key chemical composition planes with those of dwarf satellites of the MW.  The latter include the Sagittarius dSph, which is currently in the process of being engulfed by the Milky Way, as well as gas rich, irregular dwarf galaxies, such as the Large and Small Magellanic clouds (LMC \& SMC); the gas-deficient, classical dwarf spheroidal galaxies (dSph), Carina, Draco, Fornax, Sculptor, Sextans, and Ursa-Minor; and the ultra-faint dwarf spheroidal galaxy, Bo\"otes~I.  Our goal is to make inferences regarding the history of star formation and chemical enrichment of GE/S by contrasting it with MW satellites.  We note that \cite{Hasselquist2021} have performed a similar comparative analysis of GE/S and massive satellites of the MW.  In this paper we extend the analysis to lower mass satellites, with a focus on the distribution of dwarf galaxy stellar populations on the [Mg/Mn]-[Al/Fe] plane and how it can be used to constrain the history of star formation of those systems. In addition, we examine the behaviour of chemical evolution models calculated for a range of relevant input parameters on that same chemical plane.

This paper is organised as follows.  Section~\ref{sec:data} presents the data and sample adopted.  In Section~\ref{sec:chemprop} we present an examination of the distribution of the stellar populations of the various systems in key chemical planes.  In Section~\ref{sec:comparisons} we contrast the chemical composition characteristics of GE/S with those of MW satellites of various masses, using chemical evolution models to guide the discussion.  Our conclusions are presented in Section~\ref{sec:conclusions}.

%\smallskip magenta Diane feedback - You tend to refer to the star formation history and early quenching of GE/S as a well established finding. I’m not convinced this is the case. I would suggest either 1) including a discussion of other attempts to characterize the GE/S enrichment history from the literature in the Introduction and refer to them again in section 4.3, 2) expanding your discussion of the GE/S chemical evolution model parameters and how those are constrained, or 3) softening your language when discussing the enrichment history of GE/S to be a bit more speculative. }%

% ****** FIGURE ********
\begin{figure}
\includegraphics[width=\columnwidth]{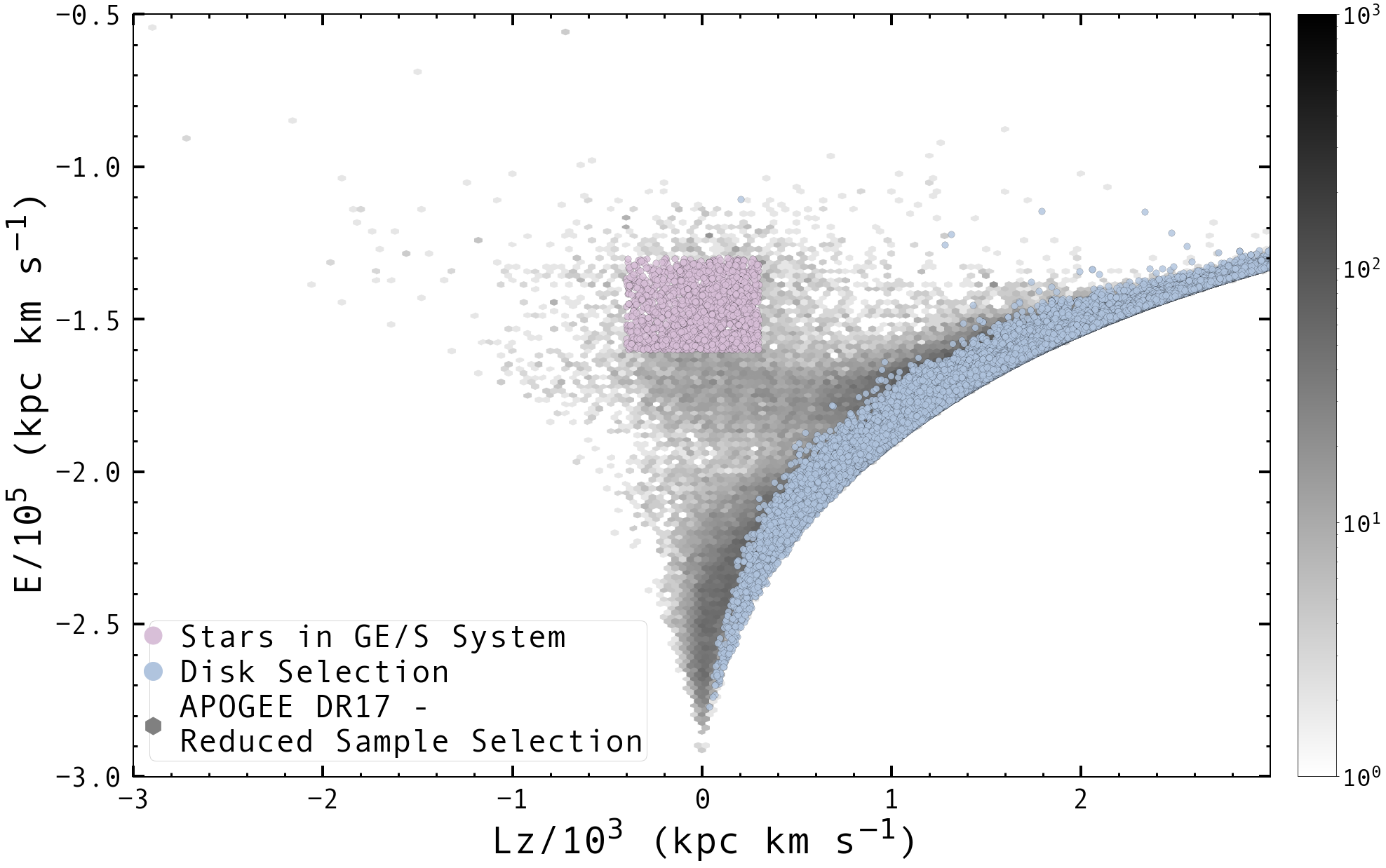}
\caption{Selecting GE/S candidates in IoM in the plane of APOGEE DR17 stars, the number density is represented by the grey-scale colour bar and does not include the following stellar populations; thin-thick disk selection ($L_{\rm Z} >$ 0, eccentricity $<$ 0.3), globular clusters [VAC by Schiavon et al. (2021, in prep)], the large and small Magellanic clouds, Sagittarius dSph and low-mass dSph galaxies. The GE/S stars in this selection (pink thistle circles) are within −1.6 $<$ E/10\textsuperscript{5} $<$ −1.3 km\textsuperscript{2} s\textsuperscript{-2} and -0.4 $<$ $L_{\rm z}$/10\textsuperscript{3} $<$ 0.3 kpc km s\textsuperscript{-1}. The selection criteria of disk stars (light steel blue circles) in this analysis have been placed on top of the plane for reference. 
\label{fig:GES_1}}
\end{figure}

\section{DATA AND SAMPLE} \label{sec:data}

\subsection{APOGEE DR17}

The second generation Apache Point Observatory Galactic Evolution Experiment (APOGEE-2, \citealt{majewski2017apache}) is part of the Sloan Digital Sky Survey IV (\citealt{blanton2017sloan}).  APOGEE-2 surveys the stellar populations of the MW with high resolution (\textit{R} $\sim$ 22,500), high S/N, spectroscopy in the H--band spectral region ($\lambda$ $\sim$ 1.51--1.70 \micron), using two twin multi--fibre spectrographs (\citealt{wilson2019apache}). Less hampered by interstellar dust than optical surveys its observations cover both Northern and Southern hemispheres, based primarily on the 2.5 m Sloan Foundation telescope (\citealt{gunn20062}) at Apache Point Observatory and the 2.5 m Iréné du Pont telescope (\citealt{bowen1973optical}) at Las Campanas Observatory.

The data employed in this paper consist of chemical compositions, stellar parameters and integrals of motion obtained from the 17\textsuperscript{th} data release \citep[DR17,][Holtzman et al. in prep.]{sdss_dr17} of SDSS-IV/APOGEE-2.  Chemical compositions and stellar parameters were generated by the APOGEE Stellar Parameter and Chemical Abundance Pipeline (ASPCAP, \citealt{perez2016aspcap}), whereas the integrals of motion result from application of the {\tt galpy} package \citep{bovy2015galpy,mackereth2018fast} to 6D phase space information resulting from combination of Gaia eDR3 proper motions (\citealt{Gaia2021}), APOGEE-2 DR17 radial velocities \citep[][Holtzman et al. in prep.]{nidever2015data}, and {\tt astroNN} machine learning-based distances (\citealt{leung2019simultaneous}). Calculations were performed adopting a \cite{mcmillan2016mass} potential for the Milky Way.

% ****** TABLE ******** 
\begin{table}
\caption{Total sample selection.}
\centering
\begin{tabular}{r c c c c c  r r r r}
\hline\hline
& Stellar &  & & & &  No. of Stars &\\
& Structures &  & & & & in Sample & \\
\hline
& Disks & & & & &  193,220 &\\[0.2ex]
& LMC & & & & &   4,610 & \\[0.2ex]
& SMC & & & & &   1,660 & \\[0.2ex]
& Sgr & & & & &   291 & \\[0.2ex]
& Bo\"otes~I & & & &  & 3 & \\[0.2ex]
& Carina & & & &  &  35 & \\[0.2ex]
& Draco & & & & &  31 & \\[0.2ex]
& Fornax & & & & &   140 &\\[0.2ex]
& Sculptor & & & & &  85 & \\[0.2ex]
& Sextans & & & & &   18 & \\[0.2ex]
& Ursa Minor & & & & & 25 & \\[0.2ex]
& GE/S & & & & &  1,952 &\\[0.2ex]
\hline
\end{tabular}
\label{table:samples}
\end{table}

\subsection{SELECTION CRITERIA} \label{sec:selection}

Our goal is to examine the chemical compositions of stars belonging to 10 dwarf satellites of the Milky Way, namely LMC, SMC, Bo\"otes~I, Carina, Draco, Fornax, Sculptor, Sagittarius, Sextans and Ursa Minor, contrasting their distributions in chemical diagnostic planes with those of the  accreted system GE/S and the MW high- and low-$\alpha$ disks. The APOGEE-2 DR17 catalogue contains data for 733,901 stars, selected according to criteria extensively discussed by \cite{zasowski2013target}, \cite{zasowski2017target}, and \cite{Santana_2021}. Before selecting stars belonging to the above systems, we must apply a number of criteria to certify the quality of the data for our analysis.  
We first cleaned the sample from stars with unreliable parameters by removing all stars with STARFLAG or ASPCAPFLAG=\textbf{BAD} \citep[see definitions in][]{holtzman2015abundances}. We then limit the data to red giant stars with S/N$>$50, stellar effective temperatures ($T_{\rm eff}$) between 3750 -- 5500 K and surface gravity (log(\textit{g})) $<$ 3.0.

Next, we removed from the sample a total of 7,562 stars that are deemed to be members of globular clusters, as listed in the Value Added Catalogue by Schiavon et al. (2022, in prep.). Application of the above filters left us with a sample of 300,389 stars. The data for the objects of interest were extracted from this surviving catalogue according to the criteria described in the following sub-sections. The final sample sizes for each system are given in Table \ref{table:samples}.

\subsubsection{Magellanic Clouds}
Our sample selection for the LMC and SMC members mimics that of \cite{nidever2020lazy} and is summarised in Table~\ref{table:MCs}. We focus on the bright and faint red giant branch (RGB) stellar populations in the MCs (see their figures 3 and 5). In this way we expect to restrict our sample to stars in approximately the same evolutionary stage as those in the MW and other satellites. 

% ****** TABLE ********
\begin{table}
\caption{MCs Sample Selection. Table summaries the sky positions ($\alpha$,$\delta$), projected distance on the sky ($d_{proj}$), Gaia proper motions (PM), radial velocities (RV), and magnitudes (H) for the LMC and SMC.}
\centering
\begin{tabular}{c c}
\hline\hline
LMC & SMC \\
$\alpha$ $\delta$ : (80.893860, -69.756126) & $\alpha$ $\delta$ : (13.18667, -72.8286)\\ [0.5ex] % inserts table %heading
\hline
$d_{proj}$ $\lid$ 12 & $d_{proj}$  $\lid$ 8 \\[1.2ex]
RV $\gid$ 125  & RV $\gid$ 100\\[1.2ex] 
2.7 $\lid$ ${\alpha}$PM  $\gid$ 1 &  2.0 $\lid$ ${\alpha}$PM $\gid$ 0\\[1.2ex]
2 $\lid$ ${\delta}$PM $\gid$ -1.2 & -0.5 $\lid$ ${\delta}$PM $\gid$ -2.0\\[1.2ex]
J-K $\gid$ 0.5 & J-K $\gid$ 0.5\\[1.2ex]
12.35 $<$ H $<$ 14.6  & 12.9 $<$ H $<$ 14.7\\[0.8ex]
\hline
\end{tabular}
\label{table:MCs}
\end{table}

\subsubsection{Sgr dSph}
Our sample for the Sgr dSph stars comes from the study by \cite{hasselquist2017apogee} and was selected by the methods described in \cite{majewski2013discovery}.
Further sampling of Sgr core and stream members are covered extensively in \cite{hasselquist2019identifying} and \cite{hayes2020metallicity}\footnote{\url{http://vizier.u-strasbg.fr/viz-bin/VizieR?-source=J/ApJ/889/63}}. Table \ref{table:Sgr_dSph} summaries the selection criteria.

\subsubsection{Dwarf Spheroidal Galaxies}
APOGEE-2 has targeted a number of dwarf spheroidal galaxies.  The field placement and target selection criteria adopted are described by \cite{zasowski2017target} and \cite{Santana_2021}.  To identify dwarf spheroidal members we first selected all stars observed within the fields of each dwarf galaxy and filtered out foreground contaminants on the basis of radial velocity and surface gravity. Stars considered members are giants ($\log g < 3.0$) whose heliocentric radial velocities differ from the central values for each galaxy by less than twice its velocity dispersion.  By proceeding in this way we prioritise sample purity over completeness.  
Radial velocities and velocity dispersions of the sample dSph galaxies are taken from Table~1 from \cite{mcconnachie2020revised} and Table~4 from \cite{mcconnachie2012observed}\footnote{\url{http://www.astro.uvic.ca/~alan/Nearby\_Dwarf\_Database.html}} with the exception of Bo\"otes~I, for which values were taken from \cite{martin2007keck}. 

%http://www.astro.uvic.ca/~alan/Nearby\_Dwarf\_Database.html

\subsubsection{Gaia-Enceladus/Sausage}\label{sec:GES}
Stars from the accreted system GE/S are distributed throughout the MW and can be discriminated through a range of chemical, kinematical and/or orbital selection criteria. In order to obtain a GE/S sample devoid of chemical composition biases, we base our selection purely on integral of motion (IoM) measurements. 

%\medskip 

After removal from the sample of all stars associated to the MW satellites, the stellar populations of the accreted system GE/S are selected on the basis of their position in the energy vs. angular momentum (E-$L_{\rm z}$) plane. 

A star is considered a member of the Gaia-Enceladus system if its energy and angular momentum fall within the following intervals:
\begin{itemize}
    \item −-1.6 $<$ E/10$^{5}$ $<$ −-1.3 km$^{2}$ s$^{-2}$
    \item --0.4 $<$ $L_{\rm z}$/10$^{3}$ $<$ 0.3 kpc km s$^{-1}$
\end{itemize}

%\bigskip 

This region of E-$L_{\rm z}$ is shown in Figure~\ref{fig:GES_1}.  The above selection criteria mimic those adopted in previous work \citep[e.g.,][]{Koppelman2019,Massari2019,Feuillet2021,Horta2021evidence}.  They are designed to take advantage of the overdensity in the E-$L_{\rm Z}$ plane around $L_{\rm Z}\sim0$ and at relatively high energy, that is easily identifiable in Figure~\ref{fig:GES_1}. We deliberately adopted a relatively high lower energy limit for our GE/S selection with an eye towards minimising disk contamination.  Yet because we imposed no chemical composition cuts, we expect a small contamination of our GE/S sample by disk stars. See discussion in Section 3.1.  This contamination by disk stars is further enhanced by the existence of ``Splash'' stars, which are early disk stars whose orbits were perturbed by the collision with GE/S \citep[][]{belokurov2020biggest}.  For a discussion of the impact of selection criteria on the chemical properties of GE/S, see \cite{Horta2022}. 

%{\magenta Diane feedback - In Section 2.2.4 on GE/S selection, I think a comparison to other GE/S selections would be interesting. Did you try adjusting the En limits to minimize MW contamination? Some discussion of how you chose these limits without clustering algorithms would be of great interest, especially to me as I have thought about this quite a bit :) If you have not yet seen it, Buder+ 2021 may also be of interest.}

% The final GE/S member selection was cleaned conservatively for consistency in the comparative analysis amongst the stellar structures, retaining a population of RGB stars in the GE/S sample with log \textit{g} $<$ 3 and $T_{\rm eff}$ $<$ 5500 K. 

\subsubsection{Thin \& Thick Disk}
The stellar populations of the the MW disk are selected using orbital parameters in IoM, focusing on stars with circular, prograde orbits. We thus retain disk stars with the following criteria:

\begin{itemize} 
 \item $L_{\rm z}$ $>$ 0
 \item eccentricity $<$ 0.3
 \item S/N $>$ 70  
\end{itemize}

The eccentricity cut employed is used to select stellar populations with disc orbits.  The adopted eccentricity threshold is arbitrary. Placing the cut at, e.g., eccentricity < 0.2 or < 0.4 would cause no impact on our analysis.  The detailed choice is not critical because the disc population is used solely as a reference for comparison with the dwarf satellite data.

%\medskip

% ****** TABLE ********
\begin{table}
\caption{Sgr Sample Selection. Table summaries the sky positions ($\alpha$,$\delta$), radial velocities (RV), surface gravity (log \textit{g}), effective temperature ($T_{\rm eff}$), and signal-to-noise (S/N) selection criteria for the Sagittarius dSph galaxy core and tail. 
}

%\footnote{\url{http://vizier.u-strasbg.fr/viz-bin/VizieR?-source=J/ApJ/889/63}}

%https://cdsarc.cds.unistra.fr/viz-bin/cat/J/ApJ/889/63

\centering
\begin{tabular}{c}
\hline\hline
Sagittarius dSph Galaxy\\
$\alpha$ $\delta$ : (284, -30)\\ [0.5ex] % inserts table %heading
\hline
90 km s$^{-1}$ $<$ RV $<$ 220 km s$^{-1}$ \\[0.4ex] 
d $>$ 5 kpc\\[0.4ex]
J − K$_{0}$ $>$ 0.8\\[0.4ex]
S/N $>$ 70\\[0.4ex]
3550 K $<$ $T_{\rm eff}$ $<$ 4200 K\\[0.4ex]
log \textit{g} $<$ 4 \\[0.4ex]
\hline
\end{tabular}
\label{table:Sgr_dSph}
\end{table}

% ****** FIGURE ******** 
\begin{figure*}
\includegraphics[width=0.7\textwidth]{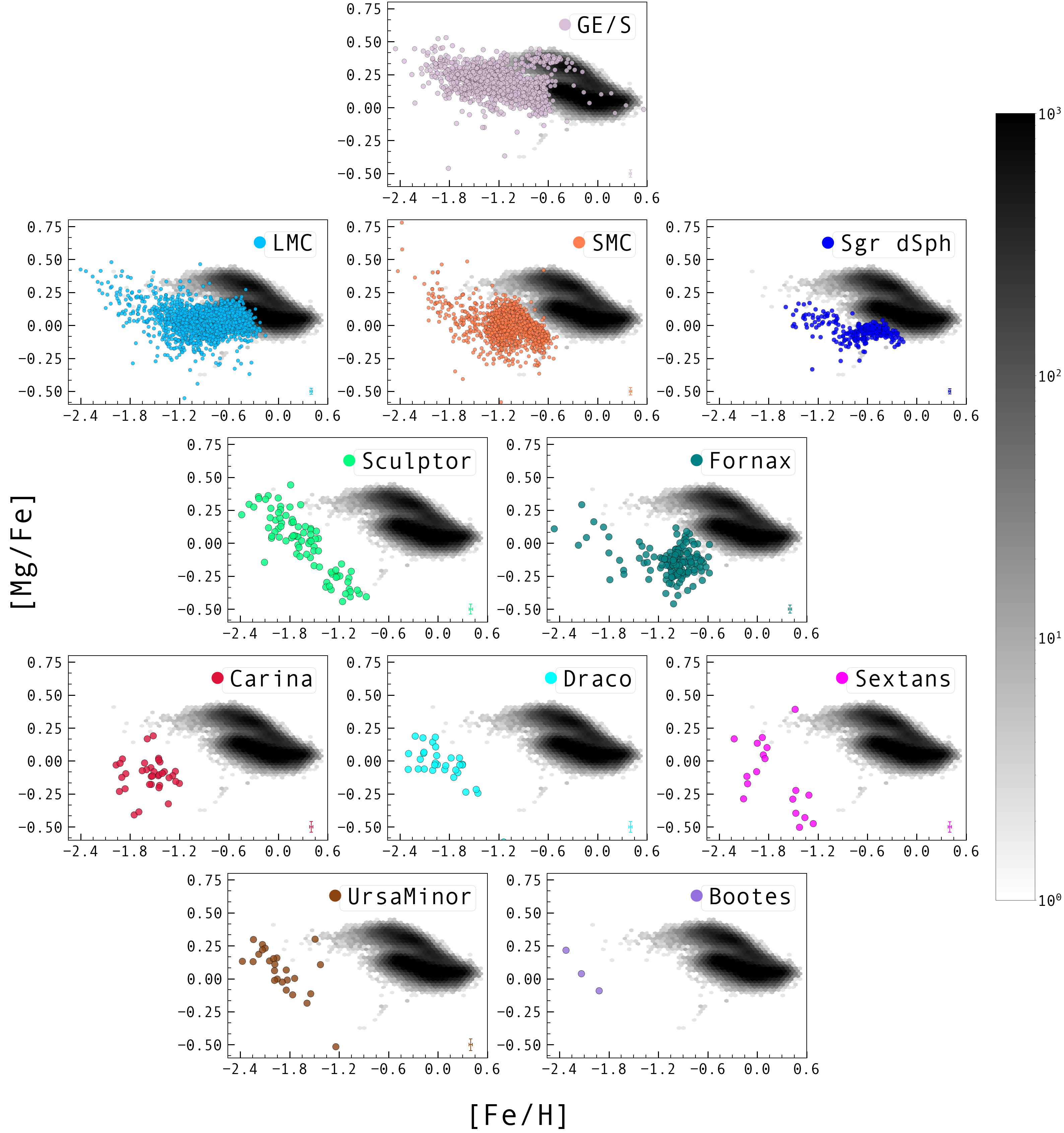}
\caption{Stellar populations of the dwarf galaxies and GE/S in the plane of MW disk (marginal density 2D hexagonal binning - the grey-scale of each hexbin denotes the number of points) in the chemical plane of [Mg/Fe] versus [Fe/H]. From top to bottom; the selected GE/S stellar population (thistle) as shown in figure 1, LMC (light blue), SMC (coral), Sgr dSph (blue), Sculptor (green), Fornax (teal), Carina (crimson), Draco (cyan), Sextans (fuchsia), Ursa Minor (brown) and Bo\"otes~I (purple). 
\label{fig:mgfe}}
\end{figure*}

% ****** Next Section ********
\section{Chemical Properties}
\label{sec:chemprop}
In the next two sections we analyse the chemical properties of the dwarf galaxies and GE/S members alongside the MW disks in the chemical planes of Mg, Mn, Al and [Fe/H].

\subsection{Magnesium}

Magnesium is an $\alpha$-element synthesised during carbon burning in massive stars, and injected into the interstellar medium during supernovae type II (SNe II) explosions (\citealt{kobayashi2006galactic}; \citealt{woosley1995evolution}). The distribution of the stellar populations on the $\alpha$-Fe plane provides important clues on the star formation history and IMF of the system (e.g., \citealt{matteucci1986}; \citealt{wheeler1989}; \citealt{mcwilliam1997}). 

%\medskip

% % ****** TABLE ********
% \begin{table}
% \caption{dSph Galaxy Sample Selection: this table summaries \textcolor{purple}{the stellar masses, (mass-to-light ratio of 1), velocity dispersions, and radial velocities for seven dwarf spheroidal galaxies}}
% \centering
% \begin{tabular}{l c c c c}
% \hline\hline
% FIELD & Stellar & Velocity & Radial & References\\
% dSph & Mass (10\textsuperscript{6}M\textsubscript{\(\odot\)}) & Dispersion & Velocity\\ [0.5ex] % inserts table %heading
% \hline
% Boötes & 0.029 & $6.5^{+2.1}_{-1.3}$ & 99.9 ± 2.4 &(1)\\[0.4ex]
% Carina & 0.38 & 6.6 ± 1.2 & 222.9 ± 0.1 &(2)(6)(7)\\[0.5ex]
% Draco & 0.29 & 9.1 ± 1.2 & -291 ± 0.1 &(2)(3)(4)\\[0.5ex]
% Fornax & 20 & 11.7 ± 0.9 & 55.3 ± 0.1 & (2)(6)(7)(9)\\[0.5ex]
% Sculptor & 2.3 &  9.2 ± 1.4 & 111.4 ± 0.12 &(2)(6)(7)(8)\\[0.5ex]
% Sextans & 0.44 & 7.9 ± 1.3 & 224.2 ± 0.1 &(2)(6)(7)\\[0.5ex]
% Ursa Minor & 0.29 & 9.5 ± 1.2 & -246.9 ± 0.1 &(2)(4)(5)\\[0.2ex]
% \hline \\ 
% \end{tabular}
% \label{table:dSph}
% {\footnotesize References: (1) \cite{martin2007keck}; (2) \cite{Grcevich&Putman}},\\ 
% {\footnotesize(3) \cite{Walker2007}; (4) \cite{Wilkinson2004ApJ...611L..21W}},\\ {\footnotesize(5) \cite{Walker2009cApJ...704.1274W};(6) \cite{Walker2009bAJ....137.3100W}},\\ {\footnotesize(7) \cite{Walker2008ApJ...688L..75W}; (8) \cite{Carignan1998AJ....116.1690C}},\\
% {\footnotesize(9) \cite{Boucher2006AJ....131.2913B}} \\
% \end{table}

% ****** TABLE ********
\begin{table}
\caption{Properties of dSph Galaxies in the sample, including ID, stellar mass, velocity dispersion, radial velocity, and original references.  All data from compilation by \protect\cite{mcconnachie2012observed}.}
\label{table:dSph}
\centering
\begin{tabular}{l c c c c}
\hline\hline
Galaxy & M$_\star$ & R.V. & $\sigma$ & References\\
 & (10$^6M_\odot$) & (km s$^{-1}$) & (km s$^{-1}$) \\ [0.5ex] % inserts table %heading
\hline
Bo\"otes~I & 0.029      & 99.9 ± 2.4     & 6.5$^{+2.1}_{-1.3}$    &(1)\\[0.4ex]
Carina & 0.38       & 222.9 ± 0.1    & 6.6 ± 1.2             &(2)(6)(7)\\[0.4ex]
Draco & 0.29        & -291 ± 0.1     & 9.1 ± 1.2             &(2)(3)(4)\\[0.4ex]
Fornax & 20         & 55.3 ± 0.1     & 11.7 ± 0.9            &(2)(6)(7)(9)\\[0.4ex]
Sculptor & 2.3      & 111.4 ± 0.12   &  9.2 ± 1.4            &(2)(6)(7)(8)\\[0.4ex]
Sextans & 0.44      & 224.2 ± 0.1    & 7.9 ± 1.3             &(2)(6)(7)\\[0.4ex]
Ursa Minor & 0.29   & -246.9 ± 0.1   & 9.5 ± 1.2             &(2)(4)(5)\\[0.2ex]
\hline \\ 
\end{tabular}
 \begin{tablenotes}[para,flushleft]\footnotesize
References: (1) \cite{martin2007keck}, (2) \cite{Grcevich&Putman}, (3) \cite{Walker2007}, (4) \cite{Wilkinson2004ApJ...611L..21W}, (5) \cite{Walker2009cApJ...704.1274W}, (6) \cite{Walker2009bAJ....137.3100W},  (7) \cite{Walker2008ApJ...688L..75W}, (8) \cite{Carignan1998AJ....116.1690C},
 (9) \cite{Boucher2006AJ....131.2913B}\\[0.5ex]
 \end{tablenotes}
 \end{table}

In Figure~\ref{fig:mgfe}, the distributions of the stellar populations in the dwarf galaxies and GE/S stars are shown in the chemical plane of [Fe/H] - [Mg/Fe] across eleven panels, each displayed alongside the stellar populations of the low- and high-$\alpha$ disks of the MW, whose chemical compositions range roughly between $\sim~-1.2~<~{\rm [Fe/H]}~<~+0.65$ and $-0.2~<~{\rm [Mg/Fe]}~<~+0.4.$  These numbers are in good agreement with independent studies based on abundance analysis of high resolution optical spectra, such as those by \cite{Aguado2021}, \cite{Matsuno2021}, and \cite{Carrillo2022}.

%\medskip

We start by pointing out that, for the reasons discussed in Section~\ref{sec:selection}, our GE/S sample contains a small, yet non-negligible, contamination by {\it in situ} stars, which can be easily spotted as they fall on the loci defined by the low- and high-$\alpha$ disks. It is fair to assume that the metal-rich stars in our GE/S sample that overlap with the high- and low-$\alpha$ disc sequences are contaminants because, on one hand, their chemical properties associate them strongly with the disc, and moreover it has been argued by \cite{Mackereth2018_origin} and Mason et al. (2022, in prep.) that low mass galaxies do not host a bimodal $\alpha$ distribution.  It is also likely that our sample may be contaminated on the metal-poor end, as {\it in situ} and accreted populations overlap on the [Mg/Fe]-[Fe/H] plane at low metallicity,  \citep[e.g.,][]{horta2020chemical}. However, the disc sequence becomes very thinly populated at [Fe/H]$\simless$--1.2, and moreover our selection criteria prioritizing high energy stars also helps minimising disk contaminations.  Thus, unless otherwise stated, our discussion henceforth ignores metal-rich contaminants overlapping with the high- and loq-$\alpha$ disk sequences on both chemical planes under study.

The latter are characterised by predominantly low [Fe/H] and lower [Mg/Fe] than high-$\alpha$ disk stars at [Fe/H]$\simgreater$--1, and overall decreasing with increasing metallicity.  On the metal-poor end ([Fe/H]$\simless$--1.8), GE/S stars reach [Mg/Fe] values comparable or even slightly higher than those of the high-$\alpha$ disk, whereas on the metal-rich end ([Fe/H]$\sim$--0.6) GE/S stars have slightly lower [Mg/Fe] than that of low-$\alpha$ disk stars of same metallicity.  The slope of the [Mg/Fe]-[Fe/H] relation undergoes a slight change, forming the so-called ``$\alpha$-knee'' at [Fe/H]$\sim$--1.2 \citep[see also][]{mackereth2019origin,Horta2021evidence}, which indicates the increased contribution of SN~Ia to the chemical enrichment of the interstellar medium.

\medskip

By and large, the locus occupied by GE/S stars in the Mg-Fe plane is somewhat similar to that where stars from the massive MW satellites LMC, SMC, and Sgr dSph are located \citep[see also][]{Hasselquist2021}.  The distributions however differ in important details.  All three of the massive satellites show, at a given [Fe/H], a positive change in the slope of the [Mg/Fe]-[Fe/H] relation, whereby [Mg/Fe] starts increasing with increasing metallicity.  This is likely associated with the occurrence of a burst of star formation in those systems, which causes an increase in the contribution of SNII to the chemical enrichment of the interstellar medium, resulting in a jump in [Mg/Fe] for increasing [Fe/H] (for a discussion, see Mason et al. 2022, in prep.).  Because star formation in GE/S was quenched at the time of accretion, no similar change in slope can be seen in its stellar populations. 

In contrast, for most low-mass dwarf spheroidal galaxies (except for Fornax), [Mg/Fe] decreases monotonically with increasing [Fe/H]. The slope of the relation is steeper than that of GE/S and the more massive satellites, and the mean [Mg/Fe] is substantially lower in low-mass satellites.  At [Fe/H]$\sim$--1.2 the stellar populations of low mass satellites are lower in [Mg/Fe] by $\sim$0.2--0.4~dex than those from their massive counterparts.

The absence of a clear ``knee'' in the $\alpha$-Fe plane of low mass satellites is likely due to it being located in those systems at metallicities that are lower than the values spanned by our sample.  Indeed, previous studies of stars in Draco, Sextans and Ursa Minor \citep{shetrone2003vlt}, Sculptor \citep{Hill2019}, and Sextans \citep{theler2020chemical} suggest the presence of a ``knee'' at [Fe/H]$\sim$--2. Based on a compilation of chemical compositions from various works \citep{CohenHuang2009,CohenHuang2010,Starkenburg2013}, \cite{Hendricks2014} quote [Fe/H]$_{knee}\sim-1.9$ for Sculptor, Ursa Minor, and Fornax and [Fe/H]$_{knee}\simless-2.5$ for Draco and Carina \citep[see also][]{deboer2014}. The low metallicity limit of our sample prohibits any statement on the existence of a change of slope for the latter two galaxies. As for the others, while our data do not rule out the presence of a change of slope in the Mg-Fe relation for Sculptor, Ursa Minor, and Fornax at [Fe/H]$\sim-2$, they cannot confirm it either, due to increased uncertainties and relatively small sample sizes in the low metallicity end.

The above result, taken together with the overall lower [Mg/Fe] of lower mass satellites, suggests that the contribution by SN~Ia to the chemical enrichment of the interstellar medium of those satellites was more dominant than in their massive counterparts, possibly indicating a lower star formation rate (SFR) throughout their histories (e.g., Mason et al. 2022, in prep.).  The one exception is Fornax, for which a sharp positive change of slope can be seen at [Fe/H]$\sim$--1.2, similar to the case of more massive satellites, suggesting also in the case of Fornax the occurrence of a burst of star formation in the recent past.

Most importantly for the goals of this study, when we contrast the position of GE/S in the Mg-Fe plane with those of dwarf satellites of the MW, we conclude that it has undergone an intense early history of star formation leading up to the build up of a relatively metal-rich and $\alpha$-enhanced stellar population, akin to those of the massive satellites of the MW.  This is not surprising, considering current estimates for the original mass of the GE/S progenitor \citep[$\sim$ 10$^8-10^9 {\rm M_\odot}$, e.g.,][]{helmi2018merger,belokurov2018co,mackereth2019origin,Deason2019,BovyMackereth2020,feuillet2020skymapper}.

\subsection{Aluminium \& Manganese [Al, Mn]}

It is well known that stellar population diagnosis is substantially improved by the consideration of the abundances of elements associated with distinct nucleosynthetic pathways.  In that vein, it has been proposed by \cite{Hawkins_2015} 
%(${\rm -1.2~< [Fe/H]}~<~-0.55$)  
and \cite{das2020ages} 
%(${\rm -2.5~<~[Fe/H]~<~-0.5}$) 
that the combination of the abundances of Fe, Mg, Mn, and Al can aid in the discrimination of accreted stellar populations in the Galactic halo.  Manganese is an Fe-peak element generated in Type Ia supernovae (SNIa) (\citealt{iwamoto1999nucleosynthesis}; \citealt{doi:10.1146/annurev.astro.38.1.191}; \citealt{2019ApJ...874..102W}). According to \cite{Hawkins_2015}, manganese is a more pure indicator of enrichment by SNe~Ia than iron, which makes that element particularly useful for chemical diagnosis. Unlike the case of [Mg/Fe], [Mn/Fe] correlates positively with metallicity (\citealt{kobayashi2020new}). This sharp distinction between the dependence of Mg and Mn on metallicity makes the ratio between these two elements a powerful discriminator between stellar populations with different chemical evolution histories.

% ****** FIGURE ********
%\begin{figure}
%\includegraphics[width=0.5\textwidth]{Images/Single_MgMnAlFe_Model.png}
%\caption{ [Al/Fe] vs. [Mg/Mn] plane colored by [Fe/H]). "Un-Evolved" and "Evolved". Figure includes stellar populations in the disk, GE/S, MCs, Sgr, Sculptor, Carina, Fornax, Draco, Ursa Minor, Sextans and Böotes, along with the chemical evolution models (see section 4.3 for a description of the models). The data is colour-coded by metallicity. 
%\label{fig:Single_MgMnAlFe}}
%\end{figure}

Aluminium, in turn, is referred to as an odd-Z element.  Although similarly to magnesium, Al is produced predominantly by SNe~II (\citealt{buchmann1984abundance}; \citealt{prantzos1996radioactive}), it can be contributed relevantly by a number of other nucleosynthetic sources.  Aluminium primarily forms during H burning phases in the CNO, NeNa and MgAl cycles (\citealt{samland1998modeling};  \citealt{guelin1995nucleosynthesis}).  A small amount of Al is also created in white dwarf binary collisions (\citealt{nofar1991formation}; \citealt{weiss199022}). Other sources of Al are observed from the winds of Wolf-Rayet (\citealt{limongi2006nucleosynthesis}) and AGB stars (\citealt{nomoto1984accreting}).  Furthermore, viable sources of $^{26}$Al in the ISM are thought to originate from the accretion of hydrogen-rich gas in white dwarf binaries following novae explosions. (\citealt{gehrz1998nucleosynthesis};  \citealt{kaminski2018astronomical}; \citealt{clayton198426}).

\smallskip

\cite{das2020ages} used the SDSS-III/APOGEE DR14 sample \citep{Albofathi2018,Holtzman2018apogee} to show that stars belonging to the GE/S system occupy a distinct locus in the [Al/Fe] vs. [Mg/Mn] plane, characterised by low [Al/Fe] and high [Mg/Mn].  More recently, \cite{Horta2021evidence} used chemical evolution models to show that this particular locus of chemical space is actually the home of {\it chemically unevolved} stellar populations.  In other words, any early stellar populations inhabit that region of chemical space, regardless of where they are formed.  As chemical evolution proceeds, the elemental abundances of {\it in situ} populations move away from that locus of chemical space, whereas the star formation of early accreted systems is quenched, so that chemical compositions are frozen in their early, pre-accretion state. 

%This chemical plane is re-created in figure \ref{fig:Single_MgMnAlFe} with the data sample analysed in this paper. A discussion on the chemical evolution models is had in section \ref{sec:mm}. 

% ****** FIGURE ******** %
\begin{figure*}
\includegraphics[width=0.7\textwidth]{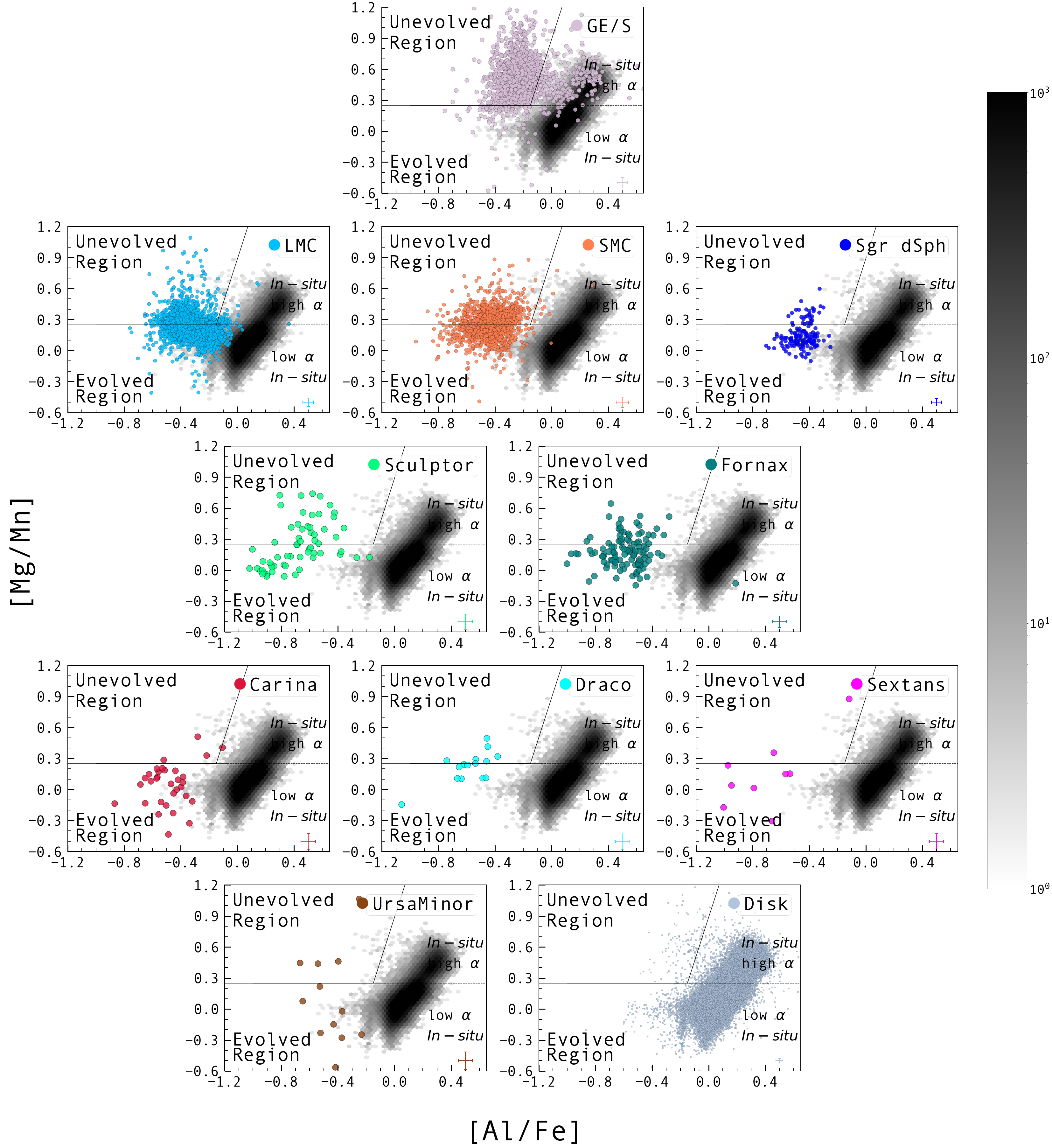}
\caption{Diagnostic plot [Al/Fe] - [Mg/Mn]. Stellar populations of the dwarf galaxies and GE/S in the plane of MW disk (marginal density 2D hexagonal binning - the grey-scale of each hexbin denotes the number of points). From top to bottom; the selected GE/S stellar population (thistle) as shown in figure 1, LMC (light blue), SMC (coral), Sgr dSph (blue), Sculptor (green), Fornax (teal), Carina (crimson), Draco (cyan), Sextans (fuchsia), Ursa Minor (brown) and the disk selection (steel blue) as shown in figure 1. Black lines separate {\it in-situ} high-$\alpha$, {\it in-situ} low-$\alpha$ stars and the unevolved region. 
\label{fig:mgmn}}
\end{figure*}

%{\magenta Diane - We have a discussion of GE/S [Mg/Mn]-[Al/Fe] and especially [Al/Fe]-[Fe/H] in Feuillet+ 2021 which you might find helpful and supportive of your work. We use a similar kinematic selection of APOGEE DR16.}

\section{Stellar Populations: Chemically Evolved or Un-Evolved}
\label{sec:comparisons}

In this section we examine the distribution of GE/S stars and MW dwarf satellites in the [Al/Fe] vs. [Mg/Mn] plane, in order to check whether the above scenario is supported by an entirely empirical examination of the data. For that purpose we compare the distribution of GE/S stars in that plane with those of the low- and high-mass satellites of the MW.  

\subsection{The detailed chemistry of a kinematically selected GE/S}  \label{sec:chemistry}

Our first goal is to check whether the locus of a GE/S sample selected purely on the basis of kinematics would be concentrated in the ``chemically unevolved'' region of the [Mg/Mn]-[Al/Fe] space.  We recall that, for this analysis to be meaningful, it is critical that the selection of the stars from all systems is entirely free of any chemical composition criterion (for details, see Section~\ref{sec:GES}).  

Inspection of Figure~\ref{fig:mgmn} shows that the majority of GE/S stars selected purely on the basis of orbital parameters fall within the ``chemically unevolved'' locus of the [Mg/Mn]-[Al/Fe] space where 82\% of GE/S stars are located.  It is important, however, to keep in mind that this line is arbitrary so that it is possible that some stars in the ``evolved'' region actually belong to GE/S. There is a small amount of contamination by high-$\alpha$ disk stars, spreading towards the high [Al/Fe] and high [Mg/Mn] (upper right) locus of the chemical space.  In contrast to the case of [Mg/Fe] abundance ratios, all the dwarf galaxies and GE/S are characterised by similarly low [Al/Fe].  In fact, at [Fe/H]$\simless$--1.0, the bulk [Al/Fe] abundances in the dwarf galaxies are lower than those in the low- and high-$\alpha$ disks.

% ******FIGURE ********

\begin{figure*}
\includegraphics[width=0.7\textwidth]{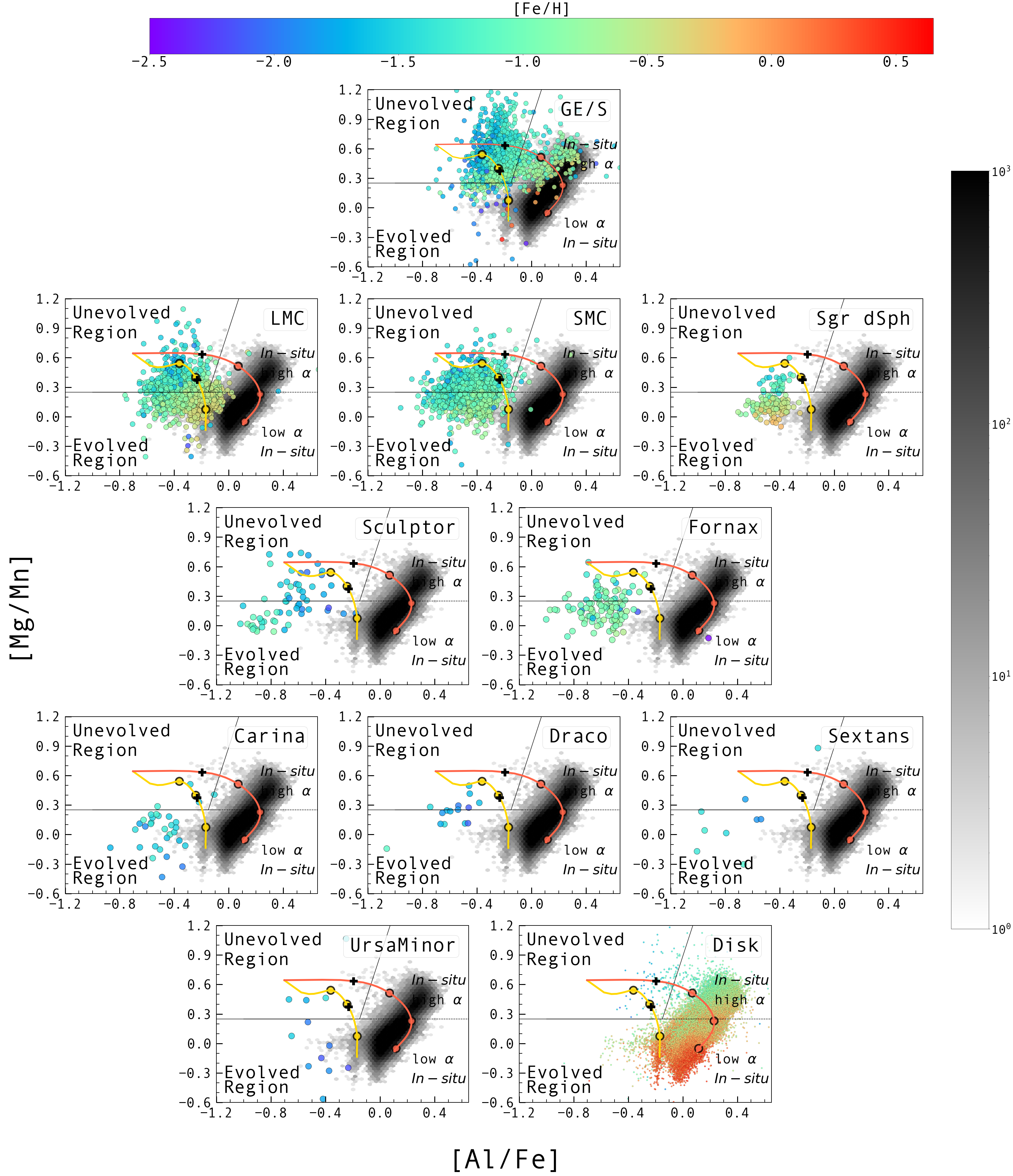}
\caption{Diagnostic plot [Al/Fe] - [Mg/Mn], colour-coded by metallicity. The MW Chemical Evolution Model (orange) and Dwarf Chemical Evolution Model (yellow) in the [Mg/Mn] vs [Al/Fe] abundance plane. Coloured circles mark the evolutionary times at 0.3, 1.0 and 5.0 Gyrs. Black cross is the position at which the models reach [Fe/H] = $\sim$-1: t= 0.1 Gyrs for the MW chemical evolution model and t = 1.2 Gyrs for the dwarf chemical evolution model. From top to bottom; GE/S, LMC, SMC, Sgr dSph, Sculptor, Fornax, Carina, Draco, Sextans, Ursa Minor and the disk. Black lines separate {\it in-situ} high-$\alpha$, {\it in-situ} low-$\alpha$ stars and the unevolved region. 
\label{fig:models}}
\end{figure*}

% ******FIGURE ********
\begin{figure*}
\includegraphics[width=0.7\textwidth]{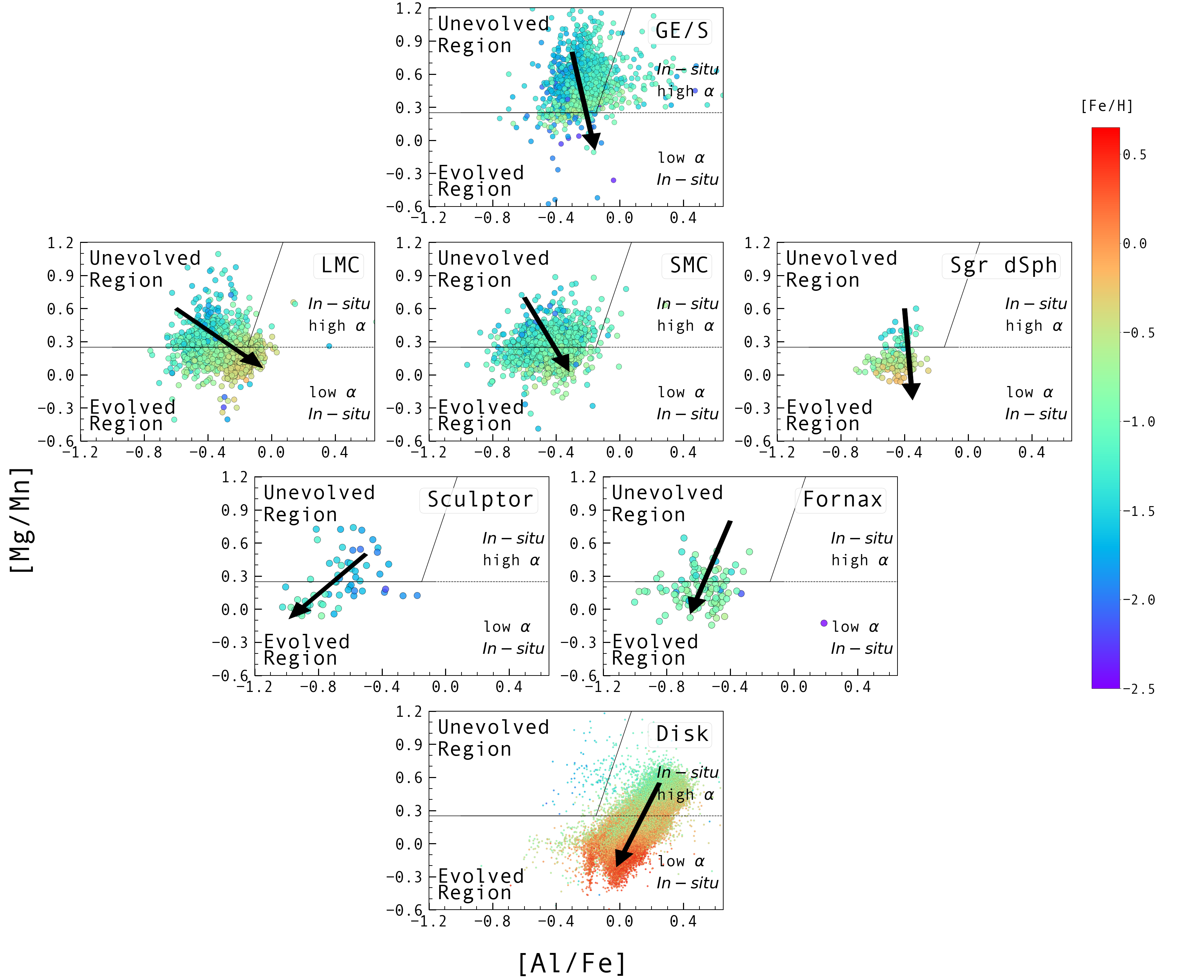}
\caption{Diagnostic plot [Al/Fe] - [Mg/Mn], colour-coded by metallicity. From top to bottom; GE/S stellar population (excluding the contaminating thick-disk/high-$\alpha$ stars), LMC, SMC, Sgr dSph, Sculptor, Fornax and the disk selection as shown in figure 1. Black lines separate {\it in-situ} high-$\alpha$, {\it in-situ} low-$\alpha$ stars and the unevolved region. In each panel, the direction of the arrow represents the metallicity vector of the stellar population. 
\label{fig:arrows}}
\end{figure*}
%{\red Laura: can you please estimate the size of this contamination?  Just please calculate the number of contaminants in the high alpha disk region over total number of GE/S stars.} \\ {\brown approximately 279 contaminates in the high alpha disk region. 279 contaminates:2600 GES stars The remaining GES sample has a mean [Mn/Mg] of  GE/S is 0.50386304. }

%{\brown The mean [Mg/Mn] of LMC is 0.23089062. The mean [Mg/Mn] of SMC is 0.26120198. The mean [Mg/Mn] of Sgr dSph is 0.14791395}

Our GE/S sample stars fall squarely within the ``chemically unevolved'' region of the plot, which we interpret as indicating an early quenching of star formation taking place during the merger of that galaxy with the MW \citep[see discussion in][]{Feuillet2021}.  This result confirms our interpretation of the distribution of the stellar populations in the [Mg/Mn]-[Al/Fe] plane, as well as the chemical evolution calculations presented in \cite{horta2020evidence}.

\smallskip

%{\magenta Diane feedback - As you do point out the large difference in star formation efficiency between the two models, a short discussion of how that was determined would be of interest.}

\subsection{GE/S vs Dwarf Satellites}
\label{sec:gesxsat}

We next compare the distribution of GE/S stars in the [Al/Fe] vs. [Mg/Mn] plane with those of the massive MW satellites (MCs and Sgr dSph).  In contrast to the case of GE/S, a substantial fraction of the stars in the MCs and the Sgr dSph lie outside that locus.  This is not surprising, as these massive satellites continued forming stars long after star formation in GE/S had ceased.  The chemically evolved, more metal-rich, stars in those massive satellites spread towards the low [Mg/Mn] region of the plane, at approximately constant [Al/Fe].  It is also worth noticing that those among the stars belonging to the massive satellite that do inhabit the ``chemically unevolved'' locus of the plot are located towards a substantially lower mean [Mg/Mn] than the stars from GE/S.  While the GE/S stars within the ``chemically unevolved'' region have mean [Mg/Mn]~$\sim$~0.50, those in the MCs and Sgr dSph have mean values between $\sim$ 0.35 and 0.25 dex lower. We speculate whether this difference is due to a selection effect caused by the fact that the massive satellites samples may be biased towards the high metallicity end of the MDF.

%{\red Add numbers.  Calculate the mean Mg/Mn of GE/S, MCs, and Sgr.  Have to remove contamination of GE/S by disk stars}

%\smallskip

%{\magenta
%\begin{itemize} 
%\item The mean [Mg/Mn] of LMC is %0.23089062
%\item The mean [Mg/Mn] of SMC is %0.26120198
%\item The mean [Mg/Mn] of Sgr dSph %is 0.14791395
%\item The mean [Mg/Mn] of GE/S is 0.49524403 (when selecting GES stars first; before disk stars removed)
%\item  The mean [Mg/Mn] of GE/S is 0.49646372 (when selecting GES stars second; after disk stars are removed first)
%\end{itemize}
%}

Differences between the loci occupied by GE/S and lower mass satellites are even more pronounced, particularly for Draco, Carina, Sextans and Ursa Minor, whose sample stars are predominantly located outside the ``chemically unevolved'' region.  While this difference may partly reflect selection biases, it is  qualitatively consistent with these lower mass satellites having undergone an evolutionary history characterised by a low, yet more prolonged, SFR than those of GE/S and the massive MW satellites, which would naturally lead to a stronger contribution to enrichment by SN~Ia and a consequently lower mean [Mg/Mn].  This is further discussed in Section~\ref{sec:models}.

\subsection{Comparison with Chemical Evolution Models}
\label{sec:mm}

In Figure~\ref{fig:models} we build on the diagnostic [Al/Fe]-[Mg/Mn] plane with the addition of two chemical evolution models calculated using the flexCE code from \cite{andrews2017inflow}.  The orange line shows the evolution of a model made to match the properties of the stellar populations of the solar neighbourhood,  representing an {\it in situ} population, and the yellow model line shows a chemical evolution model built to match the chemical properties of Gaia-Enceladus/Sausage, characterising the chemical evolution of a relatively massive satellite galaxy. The parameters adopted for the chemical evolution models are shown in Table \ref{table:new_models}. The {\it in situ} MW model is outlined in \cite{horta2020chemical}.  The model for GE/S was built to match the distribution of the data on the Si-Fe plane.  For details on the model parameters, see Tables 3 and 4 of \cite{Hasselquist2021}. The models evolve for approximately 13 Gyr: the filled circles on the models mark the evolutionary times at 0.3, 1 and 5 Gyr. The black cross marks the position at which the iron abundance reaches [Fe/H]=--1.  The star formation efficiencies in the two models differ by an order of magnitude, at 1.5~Gyr$^{-1}$ in the {\it in situ} case \citep{horta2020evidence} and 0.14~Gyr$^{-1}$ for the best fitting GE/S model \citep{Hasselquist2021}.  As a result, the solar neighbourhood model reaches [Fe/H]=--1 a mere 0.12 Gyr after the beginning of the evolution, whereas the GE/S model takes $\sim$ 1.18 Gyr to reach the same metallicity.  The models provide a qualitatively good description of the data for the {\it in situ}, accreted, and satellite stellar populations on the [Mg/Mn] vs. [Al/Fe] plane.  
These model calculations are an important tool for the interpretation of the data. In both cases, the early chemical enrichment drives the evolution towards the right due to the contribution by Type II/core collapse supernovae, because [Mg/Mn] remains approximately constant while [Al/Fe] grows due to the metallicity dependence of Al yields. By the same token, downward/left evolution on this plane reflects the increased contribution by SN Ia.

\begin{table*}
\caption{Summary of parameters used in the chemical evolution models.}
\centering
\begin{tabular}{l c c c c c}
\hline\hline
Model & MW & GE/S & LMC-like & Sgr dSph-like & Sculptor-like \\
Parameters &  &  & & & \\ % inserts table %heading
\hline
Initial Gas Mass & $2\times10^{10}\rm{M}_{\odot}$ & $3\times10^{9}\rm{M}_{\odot}$ & $3\times10^{9}\rm{M}_{\odot}$ & $3\times10^{9}\rm{M}_{\odot}$ & $3\times10^{9}\rm{M}_{\odot}$

\\
Inflow Mass Scale & $3.5\times10^{11}\rm{M}_{\odot}$ & $6\times10^{10}\rm{M}_{\odot}$ & $6\times10^{10}\rm{M}_{\odot}$ & $6\times10^{10}\rm{M}_{\odot}$ &$6\times10^{10}\rm{M}_{\odot}$  

\\
Outflow Mass Loading Factor & 2.5 & 6 & 5.4 & 20 & 40
\\
Star Formation Efficiency & $1\times10^{-9}\rm{yr}^{-1}$ & $1.5\times10^{-10}\rm{yr}^{-1}$ & $1.25\times10^{-11}\rm{yr}^{-1}$ & $2.5\times10^{-11}\rm{yr}^{-1}$ & $1\times10^{-11}\rm{yr}^{-1}$
\\
Exponential Inflow Timescale & 6 Gyr & 2.5 Gyr & 2.5 Gyr & 2.5 Gyr & 2.5 Gyr \\
\hline
\end{tabular}
\label{table:new_models}
\end{table*}

%The model for the solar neighbourhood is overall a roughly good match to the high- and low-$\alpha$ {\it in situ} populations.

The data for the various accreted/satellite stellar populations in Figure~\ref{fig:models} are colour-coded by metallicity.  One can immediately notice a difference between GE/S and all the MW satellites, which have lower [Al/Fe] on average than GE/S.  Indeed, the model is a good match to GE/S, by construction, while failing to reproduce the main locus of satellites, which is particularly noteworthy in [Al/Fe].  Given the clear dependence of the value of [Al/Fe] on star formation efficiency indicated by the models, this difference suggests that GE/S formed stars more vigorously early in its history than the MW satellites. In Section~\ref{sec:models} we discuss the behaviour of chemical evolution models on this plane in more detail.

\subsection{Metallicity gradients on the [Mg/Mn]-[Al/Fe] plane}
\label{sec:gradients}

We call attention to an important feature in the distribution of the data for different stellar populations in Figure~\ref{fig:models}.  As the  points are colour-coded by metallicity, one can see that the colour gradient on the [Mg/Mn] vs [Al/Fe] plane, varies widely from galaxy to galaxy, with more metal-rich stars being distributed towards the lower right in more massive systems and towards the lower left in low mass satellites.  This suggests that GE/S and the MCs seem to have evolved more quickly in [Al/Fe] than all the other surviving satellites of the MW (this is true even after correction for contamination by {\it in situ} stars, see Section~\ref{sec:gradients}).

We quantified this effect by measuring the metallicity gradients of our sample galaxies in the [Mg/Mn]-[Al/Fe] plane as follows.  We first reduced the sample in two ways.  For GE/S, we minimised contamination by {\it in situ} stars by selecting them in the Mg-Fe plane in the same way as \cite[][see their appendix]{Horta2021evidence}. In addition, we reduced the overall sample to stars with the most reliable elemental abundances.  The APOGEE/ASPCAP abundance measurements for Fe, Mg, and Al are quite reliable in the range of metallicities spanned by our data.  However, the abundance of Mn relies on a few relatively weak lines, becoming more uncertain in the low metallicity end.  We examined the spectra of sample stars in a range of metallicities and S/N ratios and decided to restrict the sample to stars with [Fe/H]$>$--1.6 and, for those with [Fe/H]$<$--1.5 we only kept spectra with S/N$>$70.  As a result the samples for Carina, Draco, Sextans, and Ursa Minor become too small and they are not included in this analysis.  On the basis of this reduced sample we fit linear relations to the data for each galaxy in the [Mg/Mn]-[Fe/H] and [Al/Fe]-[Fe/H] planes and derive the coefficients of the relations in the [Mg/Mn]-[Al/Fe] plane.  

The result is displayed in Figure~\ref{fig:arrows}, where arrows indicating the direction (but not the modulus) of the metallicity gradients are overlaid on the reduced samples described above. The arrows confirm the visual impression about the direction of chemical evolution on the diagnostic plane of Figures~\ref{fig:mgmn}-\ref{fig:arrows}.  In massive Milky Way satellites the arrow points towards the lower right, whereas in lower mass satellites and the Milky Way disk it points towards the lower left.  
Along the same lines, the direction of the metallicity gradient in the disk population points strongly towards the lower left of the chemical plane.  That is also the case of the disk of the Milky Way. 

%which is consistent with the fact that the MW disk has been characterised by relatively low SFR for the past several Gyr \citep[e.g.,][]{Licquia2015,RochaPinto2000}.

As indicated by the models in Figure~\ref{fig:models}, the overall direction of the metallicity gradient vector in the [Mg/Mn]--[Al/Fe] plane seems to be dependent on the star formation history of the system, which in turn is associated with its mass.  It is therefore instructive to examine the behaviour of stellar populations on this chemical plane on the basis of chemical evolution models sampling a wider range of input parameters.  That is the topic of the next Sub-section.

%The direction of the arrow is sensitive to the ratio between Mg and Al enrichment.  Thus the data suggest that the low mass satellites, along with the disk populations, are characterised by a slower enrichment in Al than Mg than the more massive satellites, where Al-enrichment follows that in Mg more closely.   

%{\cyan   In massive systems with high SFRs the early evolution of the gas moves its abundances horizontally and to the right on this plane---i.e., towards increasing [Al/Fe] and nearly constant [Mg/Mn], as indicated by the orange line.  Less massive systems and/or those with a lower SFR evolve towards lower [Mg/Mn] and, depending on the mass, slowly varying, nearly constant, or even decreasing [Al/Fe].  }

%Interestingly, GE/S is the system whose evolution seems to mimic that of the solar neighbourhood the closest, modulo the early quenching of star formation, which causes an abrupt cessation of the chemical evolution of the gas.

%{\cyan These differences between GE/S and massive MW satellites on one hand, and the low mass satellites on the other, are a likely consequence of the fact that our data for the former sample populations formed under a higher SFR. In fact, \cite{Hasselquist2021} have shown that, in order to match the distribution of the stellar populations in the $\alpha$-Fe plane of GE/S and massive MW satellites, one has to invoke high SFRs.}  

%{\purple Here goes the figure illustrating the [Mg/Mn]-[Al/Fe] planes for some chemical models where we expand the range of $\nu$ and SFE values we sample.}

\subsection{Chemical Evolution on the [Mg/Mn]-[Al/Fe] plane} \label{sec:models}

In this Section we further examine the hypothesis that the metallicity gradient in the [Mg/Mn]-[Al/Fe] plane is sensitive to the SFR.  For that purpose we calculated further chemical evolution models building on those presented in Figure~\ref{fig:models}, based on flexCE \citep{andrews2017inflow}. The same yields and SNe~Ia delay time distribution are adopted in all calculations. 
In order to produce a spectrum of models that replicate the behaviour of dwarfs of different masses in this chemical plane, we adopt a range of values for the star formation efficiency (SFE) and the wind mass-loading factor ($\eta$), which are respectively positively and negatively correlated with galaxy mass. For GE/S and MW we adopt the same models as discussed in Section~\ref{sec:mm}. Table \ref{table:new_models} summarises how these parameters differ between the models.  They are chosen to cover, in a qualitative fashion, the range of properties characteristic of the dSph galaxies included in our sample. Specific parameters adopted are tuned to approximately match the properties of the MCs (MC-like), Sgr dSph (Sgr-like) and the lower mass galaxies (Sculptor-like).  

\begin{figure}
\includegraphics{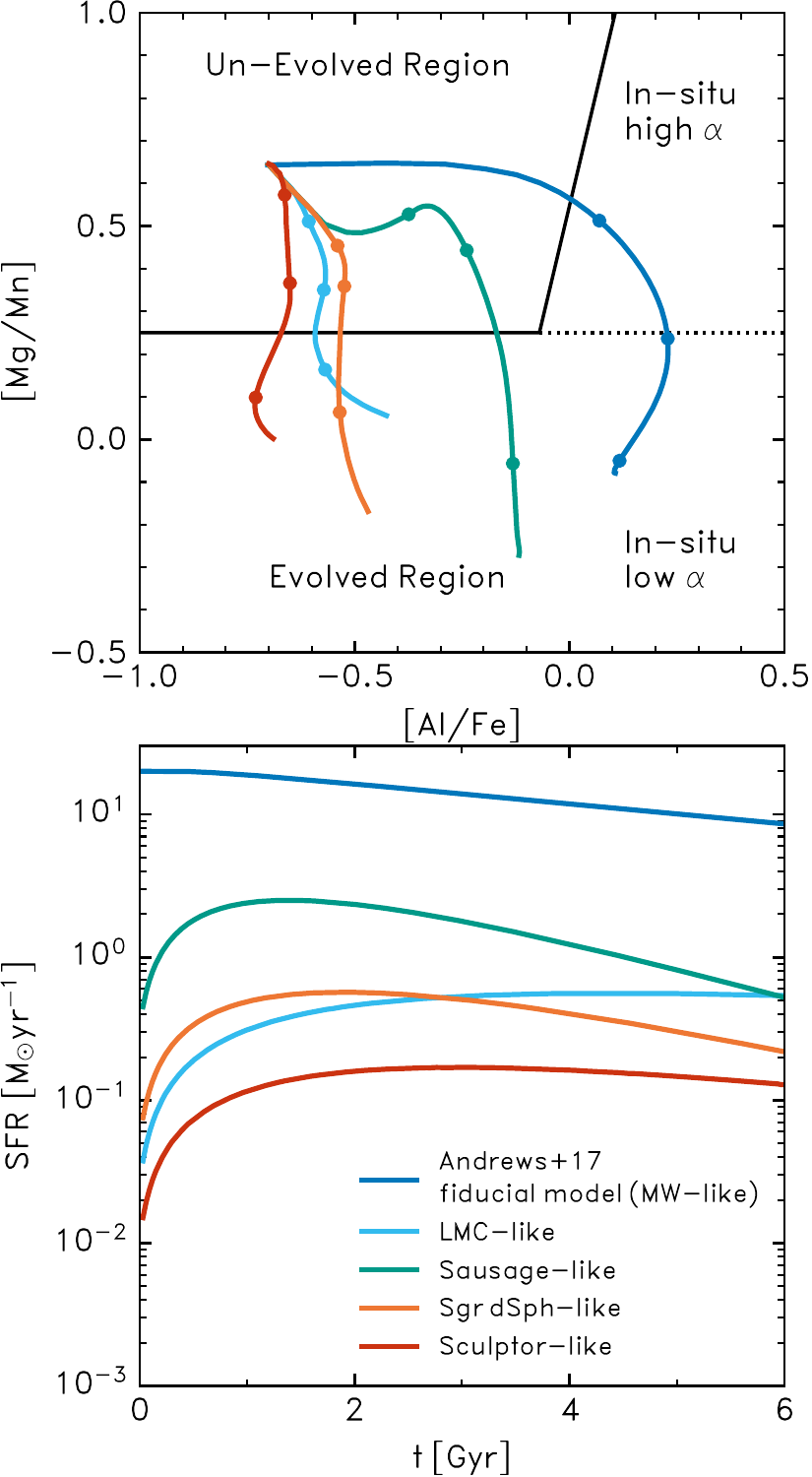}
\caption{Chemical evolution and star formation histories for a series of one-zone open box models ran using the $\textsc{flexCE}$ code. {\it Top panel:} tracks of [Mg/Mn](t) plotted as a function of [Al/Fe](t). {\it Bottom panel:} star formation rate as a function of time, SFR(t). The solid blue tracks illustrate the fiducial model of \protect\cite{andrews2017inflow} as shown in Fig.~\ref{fig:models}. For the other models, for a fixed history of gas inflow we vary the star formation efficiency and the wind mass loading factor ($\eta$)  so as to qualitatively capture the behaviour of a selection of the Local Group dwarfs in our sample with varying stellar masses.  Parameters adopted are listed in Table~\ref{table:new_models}.}
\label{fig:newmodels}
\end{figure}

The results are displayed in Figure~\ref{fig:newmodels}.  The top panel shows evolutionary tracks on the [Mg/Mn]-[Al/Fe] plane, whereas the star formation histories are plotted in the bottom panel.  As in Figure~\ref{fig:models}, evolutionary times of 0.3, 1.0, and 5.0~Gyr are indicated for each model.

\medskip

An examination of the behaviour of these various models is quite informative.  We start by considering the fiducial MW model (Table~\ref{table:new_models}), represented by the solid blue line.  It is the one for which the SFR is by far the largest, exceeding those of the other models by at least an order of magnitude.  On account of such a very high SFR, enrichment is initially entirely dominated by massive stars.  Thus the [Mg/Mn] ratio remains very high in the first few 100~Myr of evolution, whereas [Al/Fe] builds up very quickly.  
With the increasing contribution by SN~Ia at $t\simgreater$300~Myr, [Mg/Mn] starts declining steadily.  As the star formation rate declines further, so does the [Mg/Mn] ratio, and eventually the contribution by SN~Ia becomes important enough that the [Al/Fe] ratio starts declining, after about 1~Gyr of evolution.  

The remainder of this discussion contrasts the behaviour of various models within the first Gyr of cosmic evolution, where enrichment of the elements involved is dominated by CCSNe and SN~Ia. In all of the dwarf-like models, the initial SFR is substantially lower than for the MW model.  As a result, within the first few 100~Myr the evolution in [Al/Fe] is slower while [Mg/Mn] is more strongly influenced by the SN~Ia enrichment.  Hence, with decreasing SFR the early evolutionary tracks switch from pointing straight to the right to instead pointing at increasing degrees towards the lower right, and finally towards the lower left.  Indeed, as suggested by the discussion in the previous section, within the first Gyr of evolution, there is a clear correlation between the orientation of the track on the [Mg/Mn]-[Al/Fe] plane and the SFR.

This qualitative analysis can inform an interpretation of broad trends that one can promptly grasp from even a perfunctory evaluation of the data.  For instance, it is easy to see from Figures~\ref{fig:mgmn} and \ref{fig:models} that the data for GE/S are shifted towards higher [Al/Fe] and [Mg/Mn] than those of the MCs, which in turn have higher values than the lower mass MW satellites.  The numbers are summarised in Table~\ref{table:stats}.  The mean values for [Mg/Mn] and [Al/Fe] are substantially larger in GE/S than in all the MW satellites which, according to our interpretation of the models, suggests a stronger SFR in the early stages of evolution, in agreement with the results by \cite{Hasselquist_2019}.
The same conclusion can be drawn when comparing the massive MW satellites with their less massive counterparts, whose data suggest a slower rate of chemical evolution, associated with a weaker star formation rate.

\begin{table}
\caption{Mean abundances and their dispersions of MW satelites in our sample.}
\centering
\begin{tabular}{l c c c c}
\hline\hline
System & <[Al/Fe]> & $\sigma$[Al/Fe] & <[Mg/Mn]> & $\sigma$[Mg/Mn] \\
\hline
GE/S & -0.21 &   0.17 & 0.50 &  0.20  \\
GE/S (clean) &  -0.23 & 0.14 & 0.51  &   0.20  \\
SMC      & -0.42   &   0.13 &  0.27 &   0.14  \\
LMC      & -0.31   &   0.12 &  0.23 &   0.13  \\
Sgr      & -0.46   &   0.08 &  0.14 &   0.10  \\
Sculptor & -0.64   &   0.20 &  0.28 &   0.31 \\
Fornax   & -0.59   &   0.17 &  0.17 &   0.18 \\
\hline
\end{tabular}
\label{table:stats}
\end{table}

Subsequent evolution, beyond $\sim$~1~Gyr seems to be dictated by the slope of the SFR. Models with strongly decreasing SFR (e.g., MW, Sausage-, and Sgr-like) tend to evolve more strongly towards lower [Mg/Mn] with more or less constant [Al/Fe], whereas those with more approximately constant SFR (e.g., Sculptor- and LMC-like) display a slight turn over of [Al/Fe] and slower decline in [Mg/Mn].  These trends ultimately reflect the balance between the contribution by CCSNe and SN~Ia (and AGB stars in the case of Al) to the chemical enrichment of the interstellar medium (see Mason et al. 2022, in prep., for a detailed discussion).

The contrast between the histories of star formation of GE/S and massive MW satellites, and in particular the LMC and SMC is interesting in light of the fact that the mass of GE/S, according to various studies, is of the order of a few to several times 10$^8~M_\odot$ \citep[e.g.,][]{Lane2021,Deason2019,mackereth2019origin}, which is comparable to that of the SMC ($\sim$~5~$\times10^8~M_\odot$) and smaller than that of the LMC \citep[$\sim$~1.5$\times10^9~M_\odot$,][]{mcconnachie2012observed}.  That galaxies with similar masses have undergone such vastly different histories of star formation indicates a physical variable other than mass is at play at regulating the star formation histories of dwarf galaxies. \cite{Hasselquist2021} suggest it is the environment in which the dwarf galaxies formed and evolved.  We tackle this problem from the point of view of numerical cosmological simulations in a forthcoming paper (Mason et al. 2022, in prep.).

\smallskip

%{\brown from Diane Feuillet 2021 paper she says the following: The mass of the GSE progenitor has been estimated using several methods. Belokurov et al. (2018) used cosmological simulations to constrain the virial mass to be less than 10$^10~M_\odot$. Mackereth et al. (2019) constrained the stellar mass to be 3 − 10 × 10$^8~M_\odot$ by comparing dynamics and elemental abundances of accreted stars identified in APOGEE with EAGLE simulations. Mackereth & Bovy (2020) refined this estimate to be 3 × 10$^8~M_\odot$ for GSE using an analytical model of the Milky Way and APOGEE data. Mass estimates using mass-metallicity relations (Feuillet et al. 2020; Kruijssen et al. 2020; Naidu et al. 2020) or chemical evolution modeling (Helmi et al. 2018; Fernández-Alvar et al. 2018; Vincenzo et al. 2019) range from 4×10$^8~M_\odot$ to 7 × 10$^9~M_\odot$ for the stellar mass of GSE.}

\smallskip

\section{Conclusions}
\label{sec:conclusions}

We present a comparative study of the distribution in chemical diagnostic planes of the stellar populations of the Gaia-Enceladus/Sausage (GE/S) system and those of satellites of the Milky Way.  Our main conclusions are the following:

\begin{itemize}

\item We investigate the location on the [Mg/Mn] vs. [Al/Fe] plane of a GE/S sample defined purely on the basis of orbital properties.  When selected in this way, GE/S stars lie almost entirely within the locus of that chemical plane deemed to contain ``accreted'' populations by \citep{das2020ages,Hawkins_2015}.  While this result validates previous use of that method for the identification of stellar populations formed {\it ex situ}, caution is recommended, since old populations formed {\it in situ} share the same locus, as shown by \cite{Horta2021evidence}.  We therefore propose adopting a ``chemically unevolved'' nomenclature when referring to that particular locus of chemical space.  

\item The stellar populations of the satellites of the Milky Way are mostly divided between the chemically evolved and unevolved loci on this plane.  The chemically evolved, more metal-rich stars are located towards the region of lower [Mg/Mn] and approximately constant, or slightly different, [Al/Fe].  

\item The distribution of GE/S stars on the [Mg/Mn]--[Al/Fe] plane differs from those of MW satellites in an important respect.  The chemical evolution of its stellar populations in this plane suggest a higher {\it early} star formation rate than MW satellites with comparable or even higher masses, as suggested by \cite{Hasselquist_2019}. 
    
\item The direction of the metallicity vector on the [Mg/Mn]--[Al/Fe] plane is an indicator of the early star formation rate of a system.  Higher mass galaxies and/or those undergoing high star formation rates evolve more quickly in [Al/Fe] than in [Mg/Mn]. The existence of this trend is suggested by the APOGEE data on the stellar populations of the systems under study, and is boldly confirmed by the predictions of analytical chemical evolution models. The ensuing interpretation of our data on MW satellites in the light of such models leads to the conclusion that the early star formation rates of these systems was strongly affected by parameters other than galaxy mass.

\end{itemize}

\smallskip

%both mass (the gas and dust content) of the dwarf galaxies and environmental factors (ram pressure stripping and tidal stripping) have an effect on the chemical evolution of these systems. 

% Did GES bring in the other dwarf galaxies? Maybe they use to orbit GES? Did GES influence Sgr 

%All the dwarf galaxies have a chemical abundance distribution that is distinct from that of the MW disks. The chemically un-evolved stellar populations in the dwarf galaxies share similar chemical abundance ratios to the accreted system GE/S and also to the stellar populations of stars that currently reside in the Galactic halo and in stream constructs. On the other hand, the chemically evolved stellar populations in the dwarf galaxies have a different abundance distribution to the population of stars in GE/S and the halo stars: this suggests that the chemically evolved stellar populations in the dwarf galaxies formed in an environment that was different to the Galactic halo. 

\section*{Acknowledgements}

% The Acknowledgements section is not numbered. Here you can thank helpful
% colleagues, acknowledge funding agencies, telescopes and facilities used etc.

% This research made use of Astropy11 a community-developed core Python package for Astronomy, (Astropy Collaboration et al. 2013; Price-Whelan et al. 2018).

Funding for the Sloan Digital Sky 
Survey IV has been provided by the 
Alfred P. Sloan Foundation, the U.S. 
Department of Energy Office of 
Science, and the Participating 
Institutions. 

SDSS-IV acknowledges support and 
resources from the Center for High 
Performance Computing  at the 
University of Utah. The SDSS 
website is www.sdss.org.

SDSS-IV is managed by the 
Astrophysical Research Consortium 
for the Participating Institutions 
of the SDSS Collaboration including 
the Brazilian Participation Group, 
the Carnegie Institution for Science, 
Carnegie Mellon University, Center for 
Astrophysics | Harvard \& 
Smithsonian, the Chilean Participation 
Group, the French Participation Group, 
Instituto de Astrof\'isica de 
Canarias, The Johns Hopkins 
University, Kavli Institute for the 
Physics and Mathematics of the 
Universe (IPMU) / University of 
Tokyo, the Korean Participation Group, 
Lawrence Berkeley National Laboratory, 
Leibniz Institut f\"ur Astrophysik 
Potsdam (AIP),  Max-Planck-Institut 
f\"ur Astronomie (MPIA Heidelberg), 
Max-Planck-Institut f\"ur 
Astrophysik (MPA Garching), 
Max-Planck-Institut f\"ur 
Extraterrestrische Physik (MPE), 
National Astronomical Observatories of 
China, New Mexico State University, 
New York University, University of 
Notre Dame, Observat\'ario 
Nacional / MCTI, The Ohio State 
University, Pennsylvania State 
University, Shanghai 
Astronomical Observatory, United 
Kingdom Participation Group, 
Universidad Nacional Aut\'onoma 
de M\'exico, University of Arizona, 
University of Colorado Boulder, 
University of Oxford, University of 
Portsmouth, University of Utah, 
University of Virginia, University 
of Washington, University of 
Wisconsin, Vanderbilt University, 
and Yale University.

This work has made use of data from the European Space Agency (ESA) mission
{\it Gaia} (\url{https://www.cosmos.esa.int/gaia}), processed by the {\it Gaia}
Data Processing and Analysis Consortium (DPAC,
\url{https://www.cosmos.esa.int/web/gaia/dpac/consortium}). Funding for the DPAC
has been provided by national institutions, in particular the institutions
participating in the {\it Gaia} Multilateral Agreement.

Software used in this research: Astropy (\citealt{price2018astropy}; \citealt{astropy_collab}), SciPy (\citealt{virtanen2020scipy}), NumPy (\citealt{oliphant2006guide}; \citealt{harris_numpy}), Matplotlib (\citealt{hunter2007matplotlib}), Galpy (\citealt{bovy2015galpy}), TOPCAT (\citealt{taylor2005topcat}), flexCE (\citealt{andrews2017inflow}).

%%%%%%%%%%%%%%%%%%%%%%%%%%%%%%%%%%%%%%%%%%%%%%%%%%
\section*{Data Availability}

This research was made possible with data from the SDSS-IV/APOGEE-2, 17\textsuperscript{th} data release and Gaia eDR3, publicly available at \url{https://www.sdss.org/dr17/irspec/spectro_data/} and \url{https://gea.esac.esa.int/archive/}, respectively. 
 
%The inclusion of a Data Availability Statement is a requirement for articles published in MNRAS. Data Availability Statements provide a standardised format for readers to understand the availability of data underlying the research results described in the article. The statement may refer to original data generated in the course of the study or to third-party data analysed in the article. The statement should describe and provide means of access, where possible, by linking to the data or providing the required accession numbers for the relevant databases or DOIs.

\newpage


\bibliographystyle{mnras}
\bibliography{Fernandes_etal_2022} % if your bibtex file is called example.bib

%\begin{thebibliography}{99}
%New bibliography using ads bibtex citations

%\end{thebibliography}

%%%%%%%%%%%%%%%%%%%%%%%%%%%%%%%%%%%%%%%%%%%%%%%%%%

%%%%%%%%%%%%%%%%% APPENDICES %%%%%%%%%%%%%%%%%%%%%

% \appendix

% \section{Some extra material}

% If you want to present additional material which would interrupt the flow of the main paper,
% it can be placed in an Appendix which appears after the list of references.

%%%%%%%%%%%%%%%%%%%%%%%%%%%%%%%%%%%%%%%%%%%%%%%%%%

% Don't change these lines
\bsp	% typesetting comment
\label{lastpage}
\end{document}